\documentclass[fleqn,usenatbib]{mnras}

\usepackage{newtxtext,newtxmath}
\usepackage[T1]{fontenc}

\DeclareRobustCommand{\VAN}[3]{#2}
\let\VANthebibliography\thebibliography
\def\thebibliography{\DeclareRobustCommand{\VAN}[3]{##3}\VANthebibliography}

\usepackage{graphicx}	% Including figure files
\usepackage{xcolor}
\usepackage{amsmath}	% Advanced maths commands
\def\M1450{M_{\rm 1450}}
\def\CHIMP{~h^{-1}{\rm~cMpc}}

\def\Msun{{\rm M}_\odot}

\title[Reconstructing temperature profiles]{Reconstructing Large-scale Temperature Profiles around $z\sim 6$ Quasars}

\author[Huanqing Chen et al.]{
Huanqing Chen,$^{1,2,3}$\thanks{E-mail: hqchen@cita.utoronto.ca}
Rupert Croft,$^{4,5}$
and Nickolay Y.\ Gnedin$^{6,1,2}$
\\
$^{1}$Department of Astronomy \& Astrophysics,
University of Chicago,
Chicago, IL 60637, USA\\
$^{2}$ Kavli Institute for Cosmological Physics, University of Chicago,
Chicago, IL 60637, USA \\
$^{3}$Canadian Institute for Theoretical Astrophysics, University of Toronto,60 St George St, Toronto, ON M5R 2M8, Canada\\
$^{4}$ McWilliams Center for Cosmology, Department of Physics, Carnegie Mellon University, Pittsburgh, PA, 15213, USA\\
$^{5}$ NSF AI Planning Institute for Physics of the Future, Carnegie Mellon University, Pittsburgh, PA 15213, USA\\
$^{6}$ Theoretical Physics Department, Fermi National Accelerator Laboratory, Batavia, IL 60510, USA
}

\date{Accepted XXX. Received YYY; in original form ZZZ}

\pubyear{2015}

\begin{document}
\label{firstpage}
\pagerange{\pageref{firstpage}--\pageref{lastpage}}
\maketitle

% Abstract of the paper
\begin{abstract}
High-redshift quasars ionize HeII into HeIII around them, heating the IGM in the process and creating large regions with elevated temperature. 
% The IGM temperature profiles around these quasars can be used to the measure the size of the HeIII zone, and thus to constrain the  quasar lifetime. NG: repetition, the said is said at the end
In this work, we demonstrate a method based on a convolutional neural network (CNN) to recover the spatial profile for $T_0$, the temperature at the mean cosmic density, in quasar proximity zones. We train the neural network with synthetic spectra drawn from a Cosmic Reionization on Computers simulation. 
%The neural network is trained using an augmented dataset with normalized optical depth. NG: a minor detail
We discover that the simple CNN is able to recover the temperature profile with an accuracy of $\approx 1400$ K in an idealized case of negligible observational uncertainties. We test the robustness of the CNN and discover that it is robust against the uncertainties in quasar host halo mass, quasar continuum and ionizing flux. We also find that the CNN has good generality with regard to the hardness of quasar spectra. Saturated pixels pose a bigger problem for accuracy and may downgrade the accuracy to $1700$ K in the outer parts of the proximity zones. Using our method, one could distinguish whether gas is inside or outside the HeIII region created by the quasar. Because the size of the HeIII region is closely related to the total quasar lifetime, this method has great potential in constraining the quasar lifetime on $\sim $Myr timescales.
\end{abstract}

% Select between one and six entries from the list of approved keywords.
% Don't make up new ones.

%%%%%%%%%%%%%%%%%%%%%%%%%%%%%%%%%%%%%%%%%%%%%%%%%%

%%%%%%%%%%%%%%%%% BODY OF PAPER %%%%%%%%%%%%%%%%%%

\section{Introduction}

% % Lya forest and HeII heating
The temperature of the intergalactic medium (IGM) encodes the thermal history of the universe. In a broad brush picture, the IGM experienced two major heating events.
%HeII
The last one is the reionization of helium at $z\sim 3$. At this epoch, quasar activity is at its peak, and it is commonly thought that the HeII reionization is driven by quasars. HeII reionization raises the temperature of the gas by more than $\sim 10^4$ K \citep[e.g.,][]{mcquinn2009}, which can be measured from the Lyman-$\alpha$ forest \citep[e.g.,][]{schaye2000,lidz2010,becker2011,bolton2014,rorai2018,hiss2018,boera2019,telikova2019,walther2019,gaikwad2020,gaikwad2021}. 
%HI
The other heating event prior to HeII reionization is the reionization of hydrogen, commonly referred to simply as the Epoch of Reionization (EoR), which happened within the first billion years  of the universe. It is commonly thought that galaxies, instead of quasars, drive the hydrogen reionization, because the sheer number of galaxies far outweighs the number of quasars \citep[e.g.,][]{matsuoka2018}. However, because quasars outshine galaxies by orders of magnitude, in a large region surrounding a quasar called a quasar proximity zone, the properties of the IGM are heavily impacted by the quasar.

Contrary to the epoch of HeII reionization, the EoR is much less understood due to the limited direct observational data available.
Because Ly$\alpha$ absorption in the spectra of distant quasars usually saturates at $z\gtrsim 6$, it is challenging to directly measure the IGM temperature at those redshifts with the quasar spectroscopy. Therefore, most direct temperature measurements come from the quasar proximity zones, where the enhanced radiation from the quasar increases the transmitted flux and makes the IGM much more transparent. The temperature measurement in these special regions at $z\gtrsim 6$ requires not only much deeper observations to reach high signal-to-noise, but also new methodologies to accurately extract temperature information.
In the first measurements of the IGM temperature at $z\gtrsim 6$, \citet{bolton2010} fitted individual absorption lines in the proximity zones with Voigt profiles and obtained the cumulative probability distribution function (CPDF) of the Doppler parameter $b$. By fitting the measured CPDF with a suite of simulated sightlines, they constrained the IGM temperature at the mean density to be $T_0 = 23 600^{+5000}_{-6900}$ K. 
This method has been further improved and applied to a larger sample of quasar sightlines in \citet{bolton2012}.

One major motivation for directly measuring temperature at $z\gtrsim 6$ is to constrain the reionization history. However, high-$z$ quasars are thought to be located in biased regions of the universe, measurements of the IGM temperature close to a quasar may arguably be not very representative of the reionization history of the universe as a whole. On the other hand, the temperature profile can help us constrain the properties of the first quasars, in particular their lifetimes \citep{bolton2012}. The idea is straightforward: in contrast to normal galaxies, a quasar is capable of ionizing HeII to HeIII around it and, in the process, heating the IGM to temperatures above those attainable in ionization by stellar radiation. The size of such HeIII region is determined by the quasar lifetime, or the total amount of time when the central SMBH accretes in a radiatively efficient mode. By analyzing the CPDF of the Doppler parameter $b$ of seven quasars, \citet{bolton2012} has constrained the typical quasar lifetime at $z\approx 6$ to be $\gtrsim 10$ Myr.

The method adopted in \citep{bolton2012}, or the $D$-statistic, requires an accurate measurement of the distribution of $b$ parameters.
Such statistic could be improved by increasing the sample size. However, for a single sightline, the uncertainty is always $\gtrsim 4000$ K because the number of absorption ``lines'' in the proximity zone is fixed. Such uncertainty will be even bigger if we hope to obtain a temperature profile. One can expect to be able to improve the temperature profile measurement by analyzing the full spectrum instead of reducing it to a small number of fitted $b$ parameter values. This is where the convolution neural network (CNN) can come to rescue. CNNs \citep[e.g.][]{fukushima1982neocognitron,waibel1989phoneme,lecun1998gradient} have been proven powerful in picking up subtle features in a diverse set of data. Many follow-up studies have shown that CNNs have advantages over traditional methods in tasks like image classification and voice recognition, and can also be very useful in astronomy. Specifically, \citet{wang2021} and \citet{huang2021} have shown that CNN can recover the median optical depth and IGM temperature from Lyman $\alpha$ forest data with impressive precision.

In this work, we explore a method based on CNN to measure the temperature profiles in quasar proximity zones. We use the synthetic spectra from the Cosmic Reionization on Computers (CROC) simulations to exam the accuracy, robustness and generality of the CNN. Our goal here is \emph{not} to develop the complete method for the analysis of the actual observational data, such a task will require significantly more human and computational resources than that we have right now. We modestly aim at exploring theoretically the opportunities offered by CNNs and their potential reach in this problem. As we find that CNNs are indeed able to measure the gas temperature profiles around quasars in idealized environments sufficiently accurately, our work justifies pursuing this direction further in developing the pipeline for actual data analysis.

\section{Data Preparation}

\begin{figure*}
    \centering
    \includegraphics[width=0.8\textwidth]{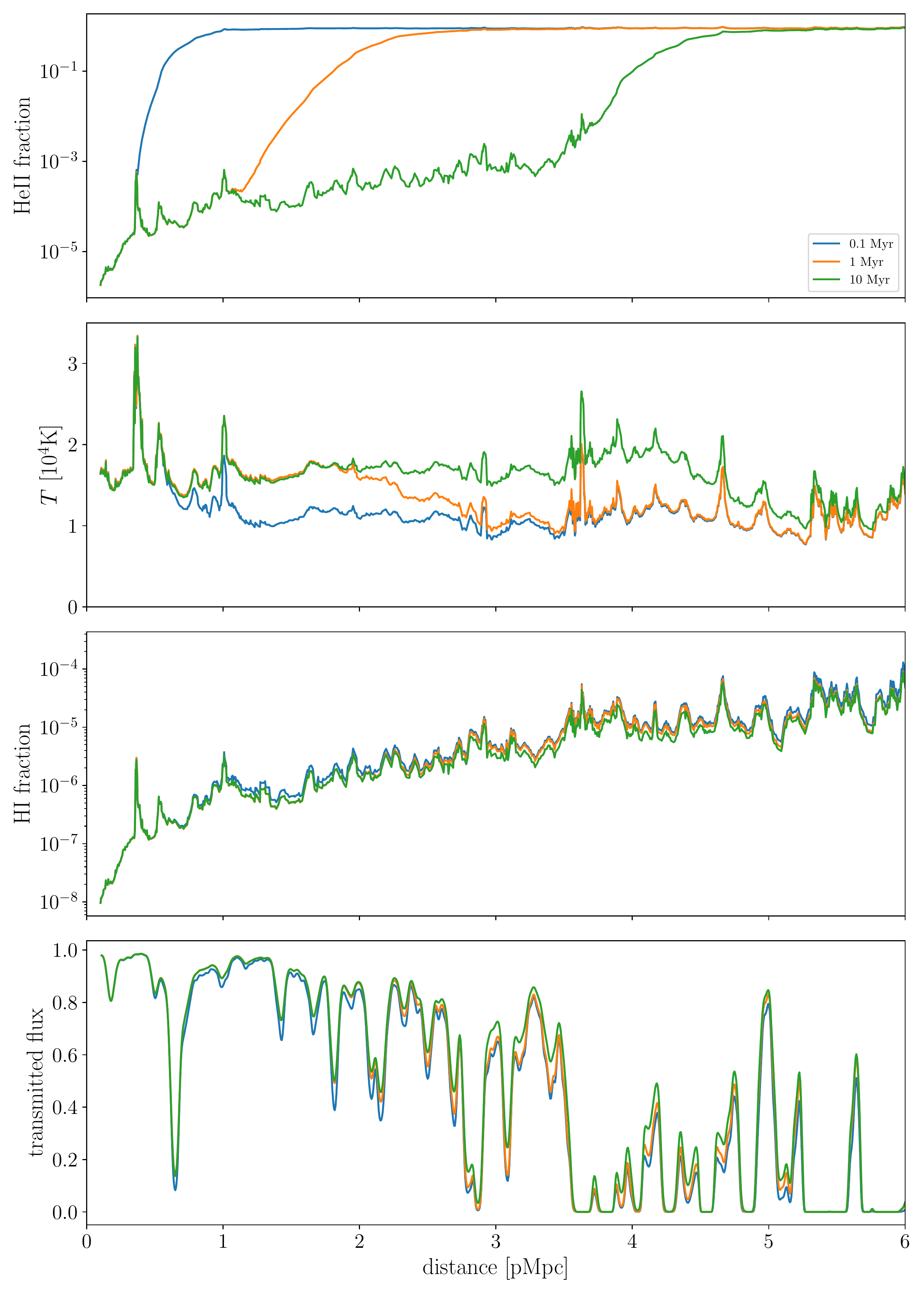}
    \caption{An example sightline from the simulation, post-processed with a quasar spectrum with the power-law index of $-1.5$. The four panels show different quantities as a function of distance from the quasar. From top to bottom: He{\small II} fraction, gas temperature, H{\small I} fraction, transmitted flux. Blue, orange and green lines show the moments when the quasar has been shining continuously for $0.1$, $1$ and $10$ Myr, respectively.
    Photoheating of He{~\small II} by the quasar significantly increases temperature. This increase in temperature slightly reduces H{~\small I} fraction and increases thermal broadening in the spectra.}
    \label{fig:exampleLOS}
\end{figure*}

\begin{figure*}
    \centering
    \includegraphics[width=\textwidth]{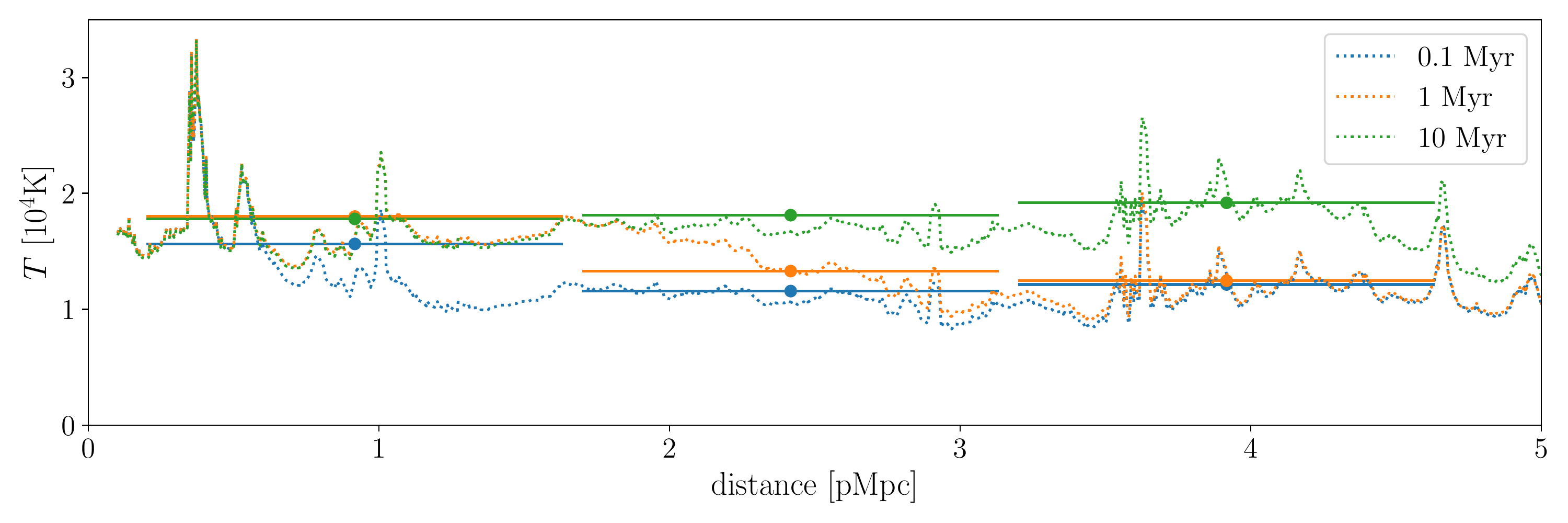}
    \includegraphics[width=1.\textwidth]{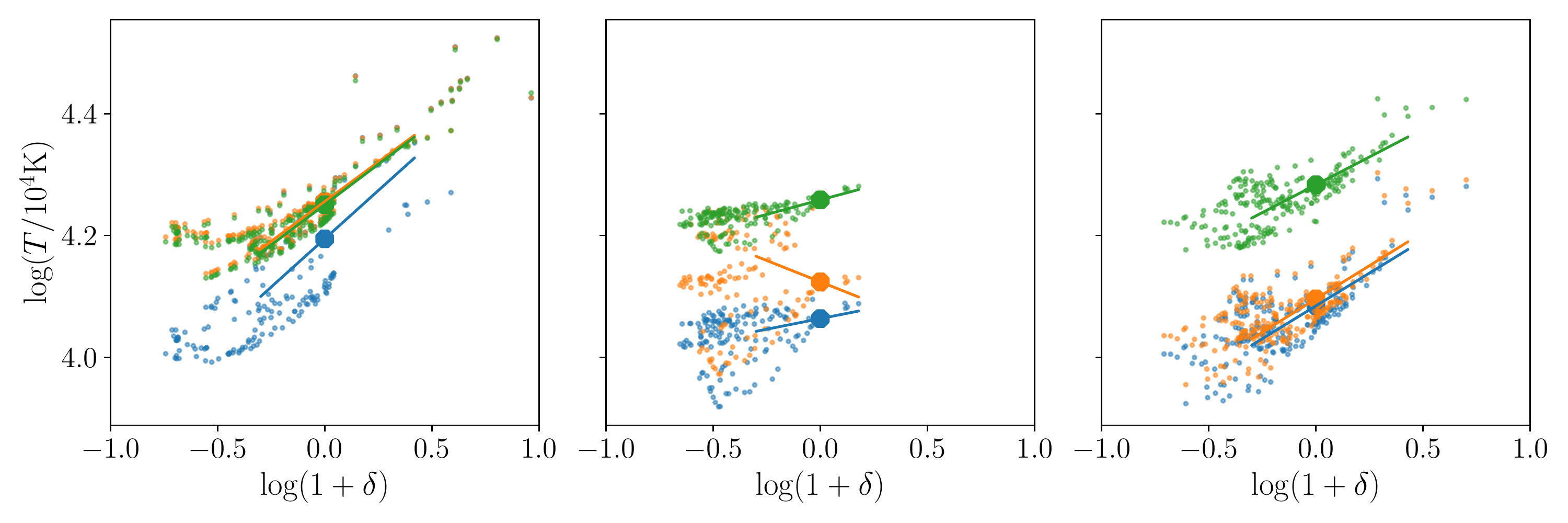}
    \caption{Top: The actual temperature profiles of the sightline (dotted lines, the same as the first 5 pMpc of the second panel in Figure \ref{fig:exampleLOS}). Dots with horizontal bars are the temperatures at the mean density of the sections covered by horizontal bars. These values are determined by fitting the power-law temperature-density relation. Bottom: the temperature-density relation, fitted by a power law. The best fit line is overlaid on the scatter plot, with the value at mean density highlighted by the large octagon. The three panels correspond to the three regions in the top panel in the horizontal order.}
    \label{fig:T_measurement}
\end{figure*}

\begin{figure*}
    \centering
    \includegraphics[width=\textwidth]{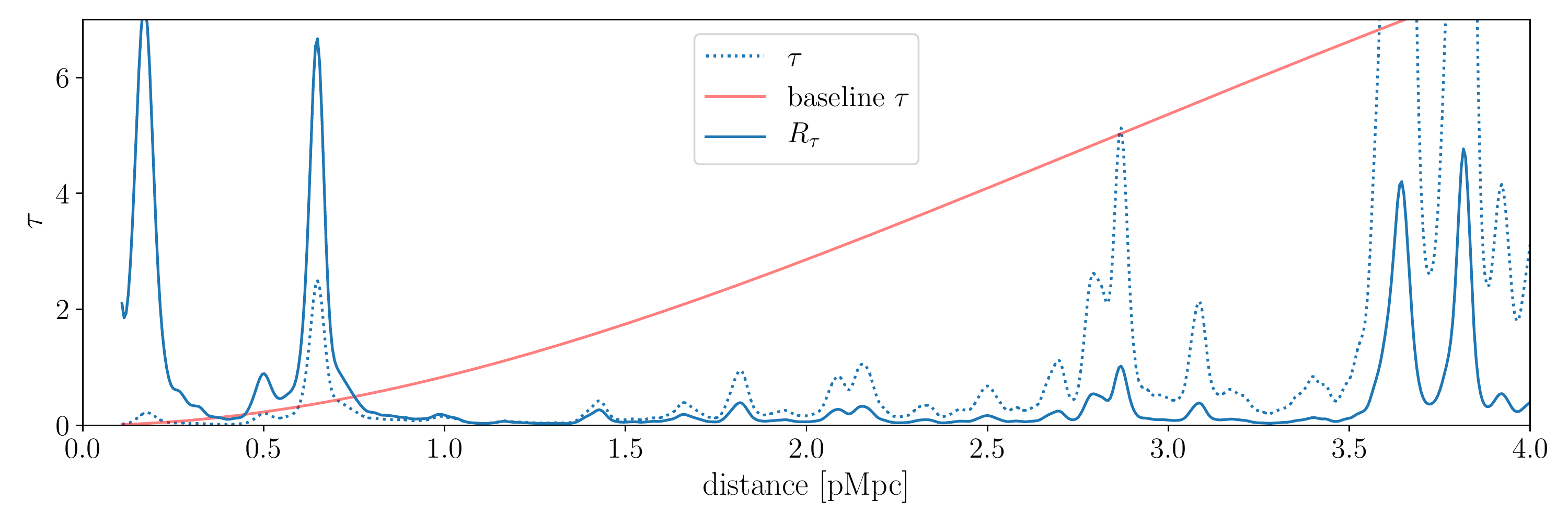}
    \includegraphics[width=\textwidth]{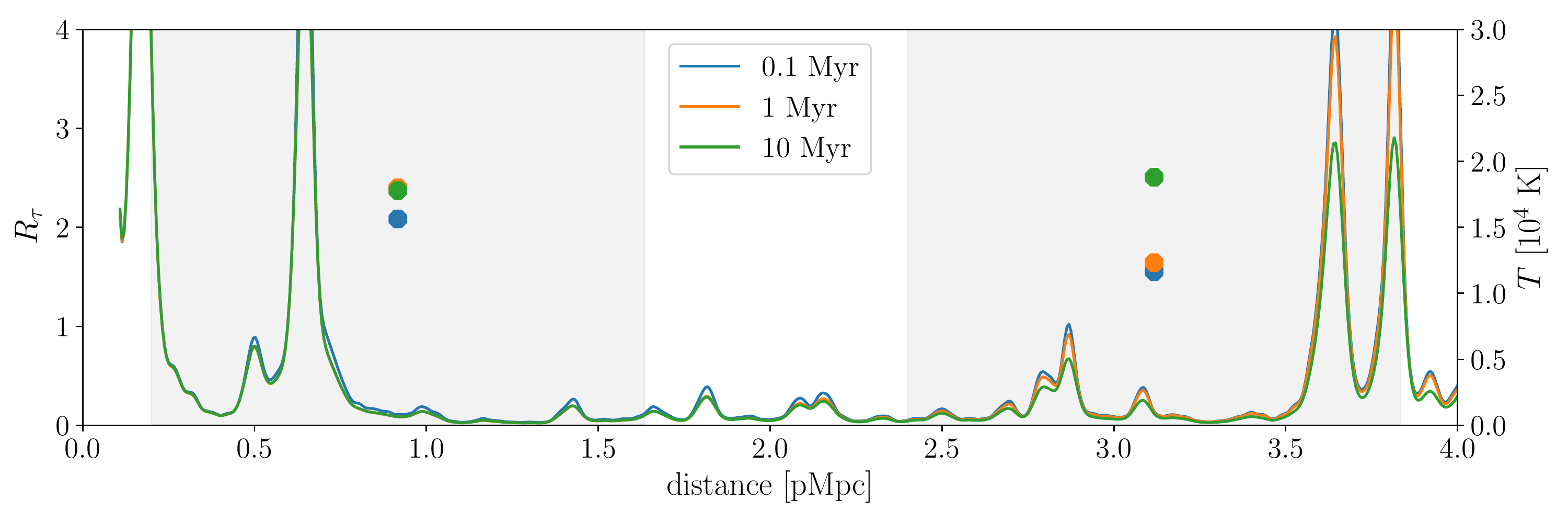}
    \caption{Upper panel: the actual optical depth of the sightline shown in Figure \ref{fig:exampleLOS} (the blue dashed line), post-processed by a quasar spectrum of index $-1.5$ for $0.1$ Myr. The red solid line is the optical depth of an imaginary sightline of uniform mean density of the universe, post-processed with the same quasar spectrum for $30$ Myr. The blue solid line shows the ``normalized'' optical depth $R_\tau$, obtained by dividing the dotted blue line by the solid red line. Lower panel: $R_\tau$ as a function of distance for different quasar lifetimes. The two grayed regions show two example data points used for our CNN training. The $R_\tau$ curves are the feature vectors, and the dots with the corresponding color are the targets (mean temperature, the right axis) we aim to predict.}
    \label{fig:training_data}
\end{figure*}

% In this study, we aim to recover the temperature profile in quasar proximity zones at $z\sim 6$.Specifically, we are interested in the profile of IGM temperature at mean density, $T_0$, defined in \ref{sec:target}. This quantity reflects the temperature on large scale and does not fluctuate much with density on small scale, and is thus useful in measuring the quasar HeIII zone and constraining the integrated quasar lifetime. In the following subsections, we show how we produce synthetic spectra from simulations and prepare the training data for the CNN. NG: this either repeats Introduction or talsk about T0 that is defined below.

\subsection{Simulations} \label{sec:simulations}

We produce proximity zone spectra from sightlines drawn from one of the CROC simulations \citep{gnedin2014} (box ``E''), similar to \citet{chen2021a}. This simulation is run with Adaptive Refinement Tree (ART) code \citep{kravtsov1997,kravtsov2002,rudd2008}. The box we use is $40 \CHIMP$ on each side, with the base grid of $1024^3$ cells, and equal number of dark matter particles, and the dark matter mass resolution of $6.2\times 10^6 \Msun$. There are additionally up to seven levels of adaptive refinement to reach the peak resolution of $100$ physical parsec that is kept constant throughout the simulation. In the simulation, stars form according to the Kennicutt-Schmidt relation \citep{schmidt1959,kennicutt1998} and emit ionizing radiation, and the radiation transfer is calculated on-the-fly using Optically Thin Variable Eddington Tensor (OTVET) \citep{gnedin2001} and is fully coupled to gas dynamics with the same spatial resolution.

We use the snapshot at redshift $z=5.9$ in this work. Similar to \citet{chen2021a}, we choose massive halos with dark matter mass $M_h \gtrsim 1.5\times 10^{11} \Msun$ as quasar hosts and draw $200$ sightlines per halo centered on them. At $z=5.9$, there are $82$ such halos in the box, resulting in $16400$ sightlines. We post-process these sightlines with a 1D RT code described in \citet{chen2021a}. We adopt a quasar emissivity of $N_{\rm ion}=1\times 10^{57} \rm s^{-1}$ and a power-law spectrum 
\begin{equation}
    L_\nu\propto \nu^{n_s}
\end{equation}
with index $n_s=-1.5$.
 We save the hydrogen neutral fraction and temperature at quasar lifetime $t_Q=0.1,0.3,1,3,10,30$ Myr and produce the Lyman $\alpha$ absorption spectra at each quasar lifetime. We call this the fiducial dataset. In Section \ref{sec:diff_spec}, in order to test the generality of the CNN, we also repeat this step with two other spectral indices $n_s=-0.5$ (a very hard spectrum) and $n_s=-2.5$ (a very soft spectrum) \citep{lusso2015}.
We then calculate the transmitted flux for each sightline. Following observational methodology, we smooth each sightline with a boxcar kernel of $700$ km/s ($20$ \AA) and measure the size of the proximity zone ($R_{\rm PZ}$), defined as the distance between the quasar and the first pixel where the smoothed transmitted flux drops below $10\%$ \citep{fan2006}. Because some sightlines encounter dense structures like Lyman $\alpha$ systems that cut the quasar proximity zone short, such sightlines are not useful for measuring the temperature profile and should be excluded from the analysis \citep{chen2021a}. Hence, we only include the sightlines with $R_{\rm PZ}>4.5$ pMpc and exclude all shorter ones. 

In Figure \ref{fig:exampleLOS} we show an example sightline. The blue, orange, and green lines in the top panel show the HeII fraction when the quasar shines continuously for $t_Q=0.1, 1,$ and $10$ Myr, respectively. Different quasar lifetime $t_Q$ results in different size of the HeIII region. HeII photo-ionization elevates gas temperature, as shown in the second panel. Hydrogen neutral fraction, which is shown in the third panel, is reduced slightly in HeIII region because recombination rate decreases at higher temperature. Higher temperature also results in larger Doppler broadening in the spectra, which is shown in the bottom panel. It is this enhanced Doppler broadening that the CNN is expected to capture.

% \section{Some extra material}

% Temperature causes thermal broadening in absorption. In the last panel of Figure \ref{fig:exampleLOS}, we can find that if the gas is further photo-heated by HeII reionization, small-scale ($\lessim 15$ km/s, equivalent to $\lessim 22$ pkpc) fluctuations in the transmitted flux are smoothed (see the region between $3-3.5$ pMpc, for example). 

\subsection{Training Data}

Our goal is to use a CNN to predict the temperature at the mean density $T_0$, the CNN "target", using spectral features. In this subsection we first describe how we define our target in Section \ref{sec:target}. We then describe how we generate the feature vector in Section \ref{sec:feature}.

% Our goal is to train a CNN, similar to \citet{wang2022}, to pick up such features to predict the mean temperature of the gas. The spectra in the proximity zones at $z\sim 6$, however, differ significantly from Lyman $\alpha$ forests at $z=2-5$ because it has a clear dependence on the distance to the quasar. Nonetheless, we can ``normalize'' it using the method in \citet{chen2021b}, which render the ``normalized'' spectra nearly independent from distance. This method can also help us significantly increase the size of training dataset. We describe how we prepare the augmented dataset in Subsection \ref{sec:feature}. 

\subsubsection{Temperature at the mean density $T_0$ : the target} \label{sec:target}

We define the temperature at the mean density $T_0$ by fitting a power law to the temperature-density relation \citep{hui1997}. In Figure \ref{fig:T_measurement}, we show the process of obtaining $T_0$ of some sections of an example sightline. The dotted lines in the upper panel show the temperature profiles of the sightline at different quasar lifetimes $t_Q=0.1$ Myr (blue), $1$ Myr (orange) and $10$ Myr (green). It has small scale features correlated with the gas density. 
%In this study we are more interested in the ``typical'' temperature of a section. NG: there is no "typical" temperature.
In order to measure $T_0$ at different distances from the quasar, we divide the whole spectrum into separate sections of $1024$ km/s in length (in velocity space along the spectrum), which corresponds to $\approx 1.4$ pMpc. Three sections are marked by the horizontal line sections in the upper panel of Figure \ref{fig:T_measurement}. We then use all cells in each section and fit a power law relation between the gas temperature and the gas density. Temperature--density relations for the three sections on the upper panel are shown in lower panels in the same order. We fit a power-law in the density range $0.5<1+\delta<3$ to avoid the pixels at the low and high density ends, which usually cannot be fit well by a simple power-law. The value at $\delta=0$ at the best-fit power-law is then defined as the temperature at the mean density $T_0$, shown as a big octagon in each of the lower panels. We also plot thus defined $T_0$ in the upper panel as circles. 
We check the robustness of our measurement of $T_0$ by changing the density range of the fit to between $0.3$ and $5$ and find that the difference is $<2\%$, showing that our measurement of $T_0$ is very robust even if the scatter in the lower panels of Figure \ref{fig:T_measurement} is not small.

In principle, one can also explore the slope of the temperature-density relation. In this work we deliberately restrict our focus on the temperature only, as the slope is likely to be a harder feature to capture with a simple CNN -- it would manifest itself in small changes in the Doppler broadening of features of different amplitude rather than the overall change in the Doppler broadening induced by $T_0$. It may be interesting follow-on direction, though.

\subsubsection{Normalized optical depth: the feature vector} \label{sec:feature}

The average transmitted flux decreases rapidly away from the quasar due to the decrease of ionizing flux. However, \citet{chen2021b} shows that because the proximity zone is so transparent, the ionizing flux is inversely proportional to the square of the distance from the quasar. We can thus ``normalize'' the optical depth by dividing it by the ``baseline'' optical depth ${\tilde{\tau}}$, which is obtained by post-processing an imaginary sightline of the uniform density at cosmic mean. In this work the baseline $\tilde{\tau}$ is chosen at quasar lifetime $t_Q=30$ Myr. We do not change it for different $t_Q$ because in reality this quantity is not known beforehand. We only need a smooth baseline to get rid of distance dependency so that we can augment the training data, so any value for $t_Q$ in excess of 30 Myr will result in identical outcome. Hereafter, we denote this normalized optical depth as $R_{\tau}$:

\begin{center}
\begin{equation}\label{eq:Rtau}
R_{\tau}=\frac{\tau}{\tilde{\tau}}
\end{equation}
\end{center}

In fact, the square root of $R_\tau$ is the thermally smoothed density field \citep{chen2021b}. In the upper panel of Figure \ref{fig:training_data} we show one such sightline.
%The dotted blue line is the optical depth $\tau$ of the sightline at quasar lifetime $t_Q=0.1$ Myr. The red line shows the baseline $\tau$. The normalized optical depth $R_\tau$ is shown by the solid blue line. NG: you don't need to repeat the caption here.
With this simple operation, the normalized optical depth is almost independent of the distance to the quasar in the proximity zone. We can thus randomly draw sections in the proximity zones to augment the training set. In practice, for each sightline, we randomly choose $10$ sections within $0.2-4.2$ pMpc, each of them $1024$ km/s long. For each section, we use an array of $256$ pixels of $\log_{10} (R_\tau)$ as the feature vector and calculate its $T_0$ as the target. We thus produce a dataset of $\approx 400,000$ target--feature pairs.  In the lower panel of Figure \ref{fig:training_data}, we show example training data chosen from two random sections, where $R_\tau$ in each grey region is used for making the feature vector and the target temperature is plotted as an octagon of the same color (the temperature axis is on the right). The three colored lines in the lower panels show the cases at different quasar lifetimes. We can see that in the outer section when the gas is further heated significantly by HeII ionization at $t_Q=10$ Myr, the normalized optical depth $R_\tau$ is smoother than when the gas is not yet heated  ($t_Q=0.1$ Myr and $1$ Myr).

% \begin{figure*}
%     \centering
%     \includegraphics[width=0.48\textwidth]{fig/histogram.pdf}
%      \includegraphics[width=0.48\textwidth]{fig/histogram_even.pdf}
%     % \includegraphics[width=1.\textwidth]{fig/fit_ave_temp_diff_spec.pdf}
%     \caption{}
%     \label{fig:hist}
% \end{figure*}

\section{The CNN model} \label{sec:cnn}

\begin{figure*}
    \centering
    \includegraphics[width=0.99\textwidth]{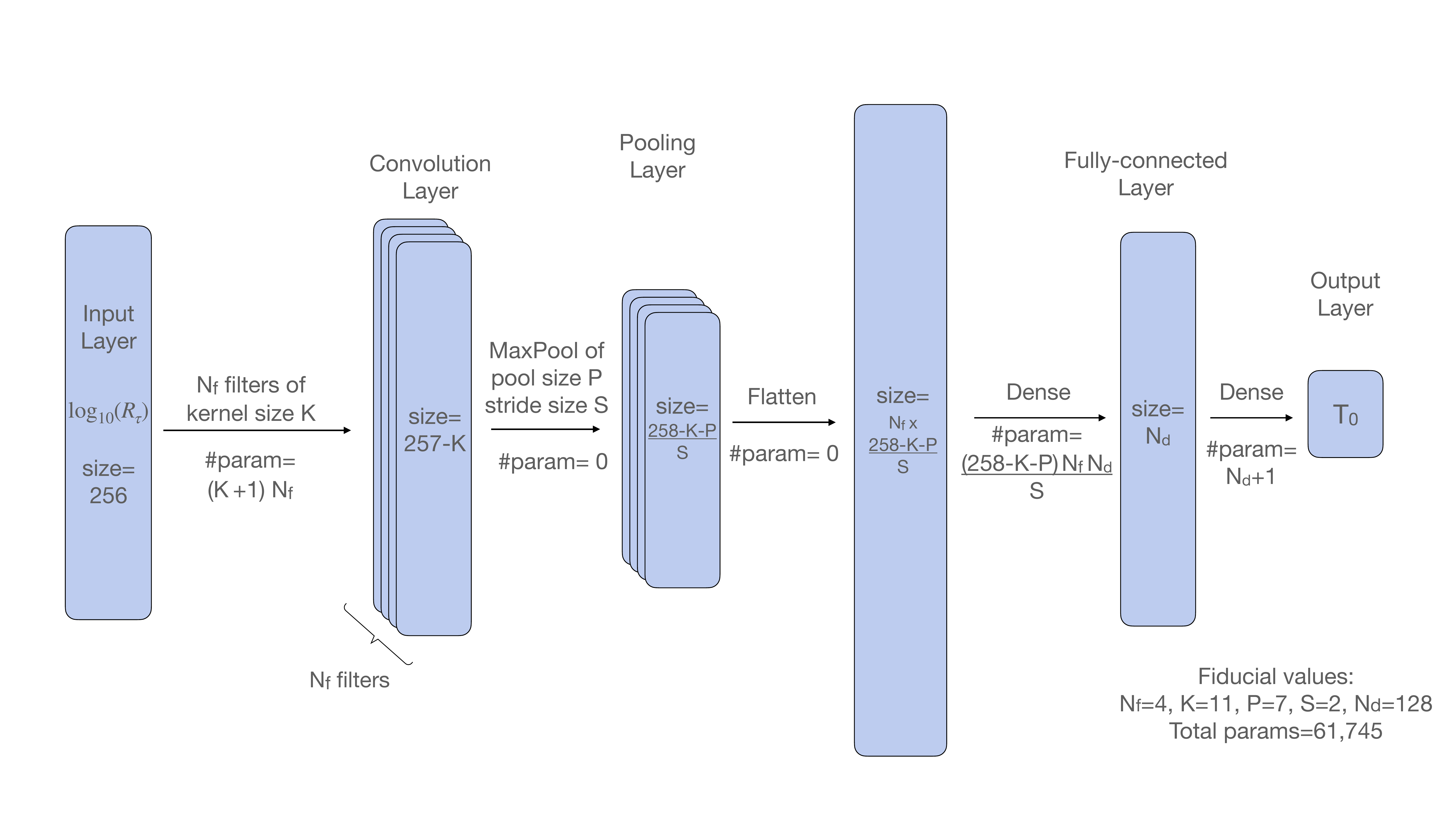}
    \caption{CNN architecture.}
    \label{fig:CNN}
\end{figure*}

% In the meantime, we explore the hyper parameter space and train using different combination of training data to test the robustness and generality of the models. 
% As shown in the following, this simple architecture works surprisingly well.

\subsection{Architecture}

In this study, we use the simplest CNN architecture (Figure \ref{fig:CNN}) to predict $T_0$. This architecture has only one convolutional layer, one pooling layer, and one fully connected layer. This simple CNN has proven to work on $z\approx 3$ Lyman $\alpha$ forest in previous works \citep{huang2021,wang2021}. In image classification with multiple convolutional layers, it has been found that the first few convolutional layers extract low-level features like edges, while the last few layers extract high-level features of complex patterns \citep[e.g.,][]{zeiler2013}. Because temperature is believed to impact only the curvature of absorption features, which is a low-level feature, we do not explore more complex CNNs with multiple convolution layers. Following \citet{wang2021}, we fix the pooling method to be the maximum and activation functions in the fully connected layers to be Scaled Exponential Linear Unit (SELU) \citep{Klambauer2017}:
\begin{equation}
    f\left(x\right) =
    \begin{cases}
    \lambda{x} & \text{ if } x \geq{0} \\
    \lambda{\alpha\left(\exp\left(x\right) -1 \right)} & \text{ if } x< 0
    \end{cases}       
\end{equation}
with $\lambda=1.05070098$ and $\alpha=1.67326324$. We use the mean squared error (MSE) as the loss function and the Adam optimizer \citep{kingma2014} for training the CNN. To train and evaluate the CNN, we split the whole data set into a training set, a validation set, and a test set with ratios of $7:2:1$, and we use a batch size of 512.

\subsection{Hyper parameter tuning and results}

Even the simplest CNN has multiple hyper parameters that need to be chosen: the number of filters $N_f$, the kernel size $K$, the pool size $P$, the stride size $S$, and the number of neurons in the hidden fully-connected layer $N_d$ (Figure \ref{fig:CNN}). We explore a grid with $N_f=(4,8,12)$, $K=(5,7,9)$, $P=(3,5,7)$, $S=(1,2,4)$, and $N_d=(128,256)$. In order to explore the large space of hyper parameters with limited GPU resources, we first run each set of hyper parameters for $500$ epochs. 
% We find that the combination of (4,9,7,2,128) achieves a small validation loss of $\sqrt{\rm MSE} \approx 1600$ K yet has the smallest number of trainable parameters.
Across these parameter combinations, we find the validation loss is not sensitive to $N_f, P, S$ and $N_d$. However, the validation loss decreases with $K$ slightly. We thus test a few more models with $K=11, 13, 15, 17, 19$. We find the combination of $(N_f,K,P,S,N_d)=(4,11,7,2,128)$ returns the smallest validation loss and we choose this combination as the fiducial hyper parameters.

% \subsection{test result}

\begin{figure}
    \centering
    \includegraphics[width=0.45\textwidth]{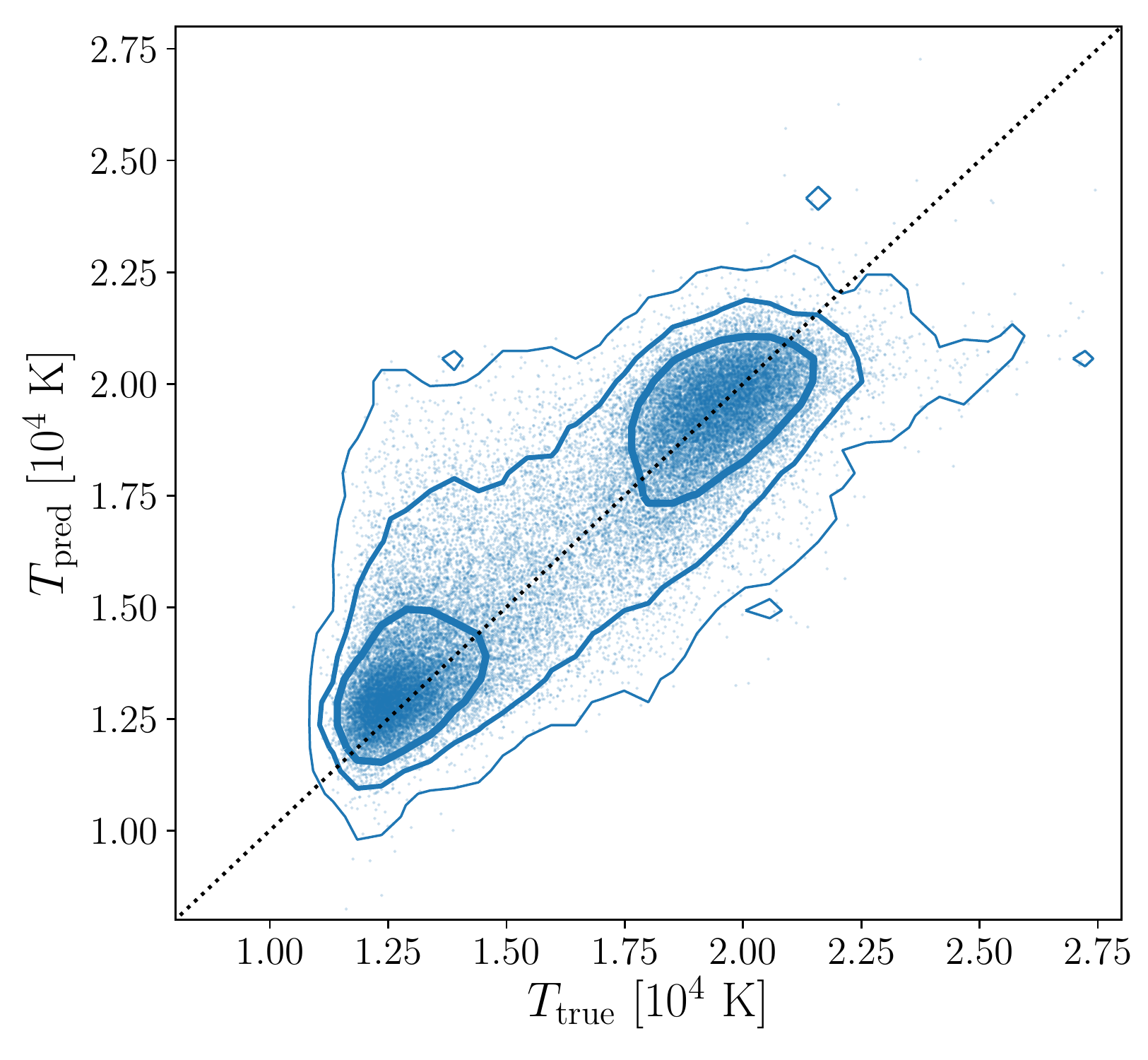}
    \caption{Scatter plot of true $T_0$ vs $T_0$ as predicted by the CNN. The contours represent $68\%$, $95\%$ and $99.7\%$ of the test data.}
    \label{fig:cnnRes}
\end{figure}

We continue to train the model for $2000$ epochs and find that $\sqrt{\rm MSE}$ of the validation set converges to $1400$ K after $\approx 1500$ epochs. We then examine the result in the test set. The $\sqrt{\rm MSE}$ of the test set is the same as the validation set. In Figure \ref{fig:cnnRes} we show the scatter plot of true $T_0$ versus predicted $T_0$ of the test set, overlaid by $68\%, 95\%$ and $99.7\%$ contours. The true temperature is bimodal, corresponding to gas inside (hotter) and outside (cooler) of the HeII I-front. The error of the  predicted $T_0$ is small enough to distinguish the two cases.

% In Figure \ref{fig:exampleTpredLOS}, we show the recovered profile of a sightline.

\begin{figure*}
    \centering
    \includegraphics[width=0.37\textwidth]{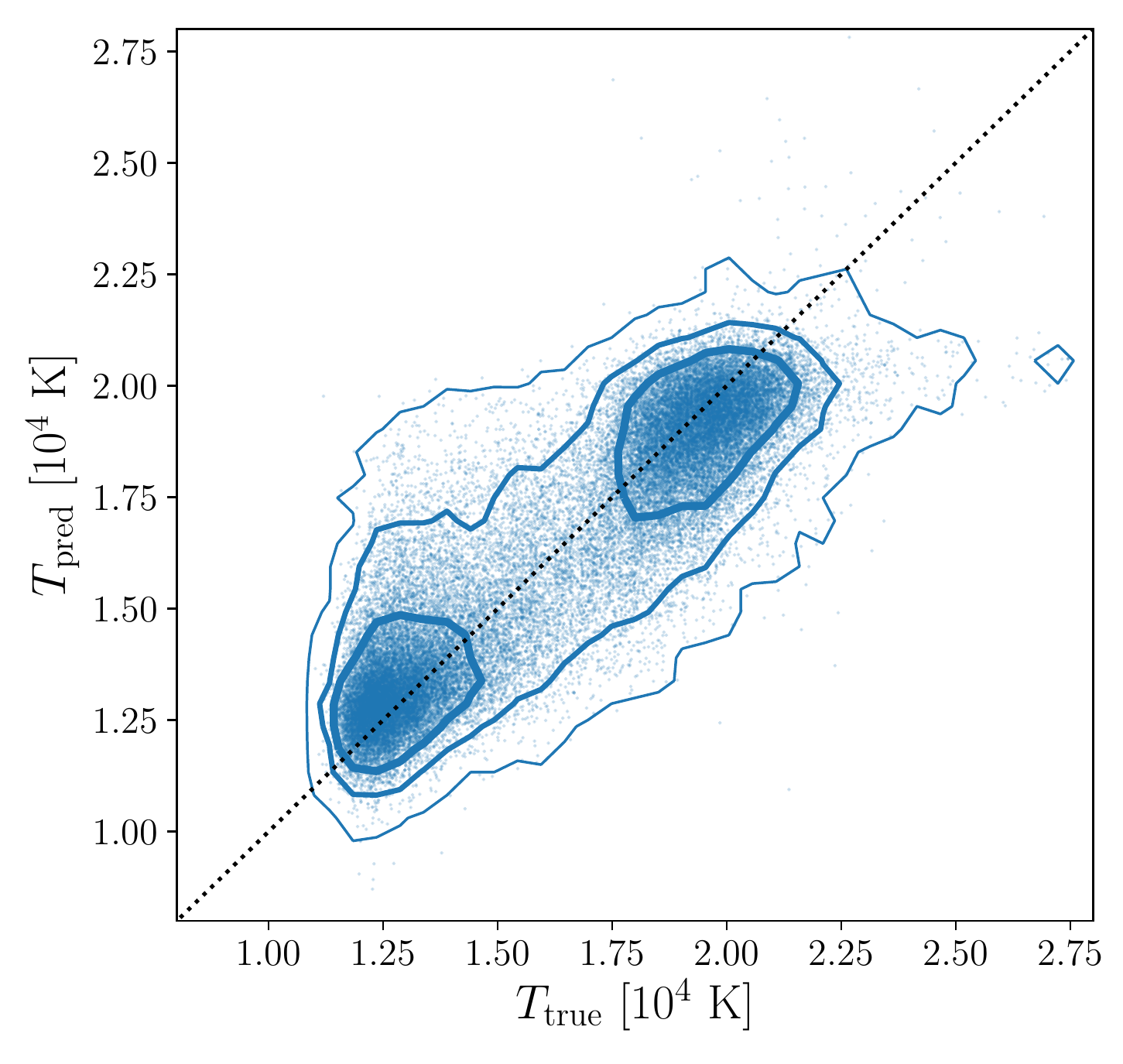}
    \includegraphics[width=0.6\textwidth]{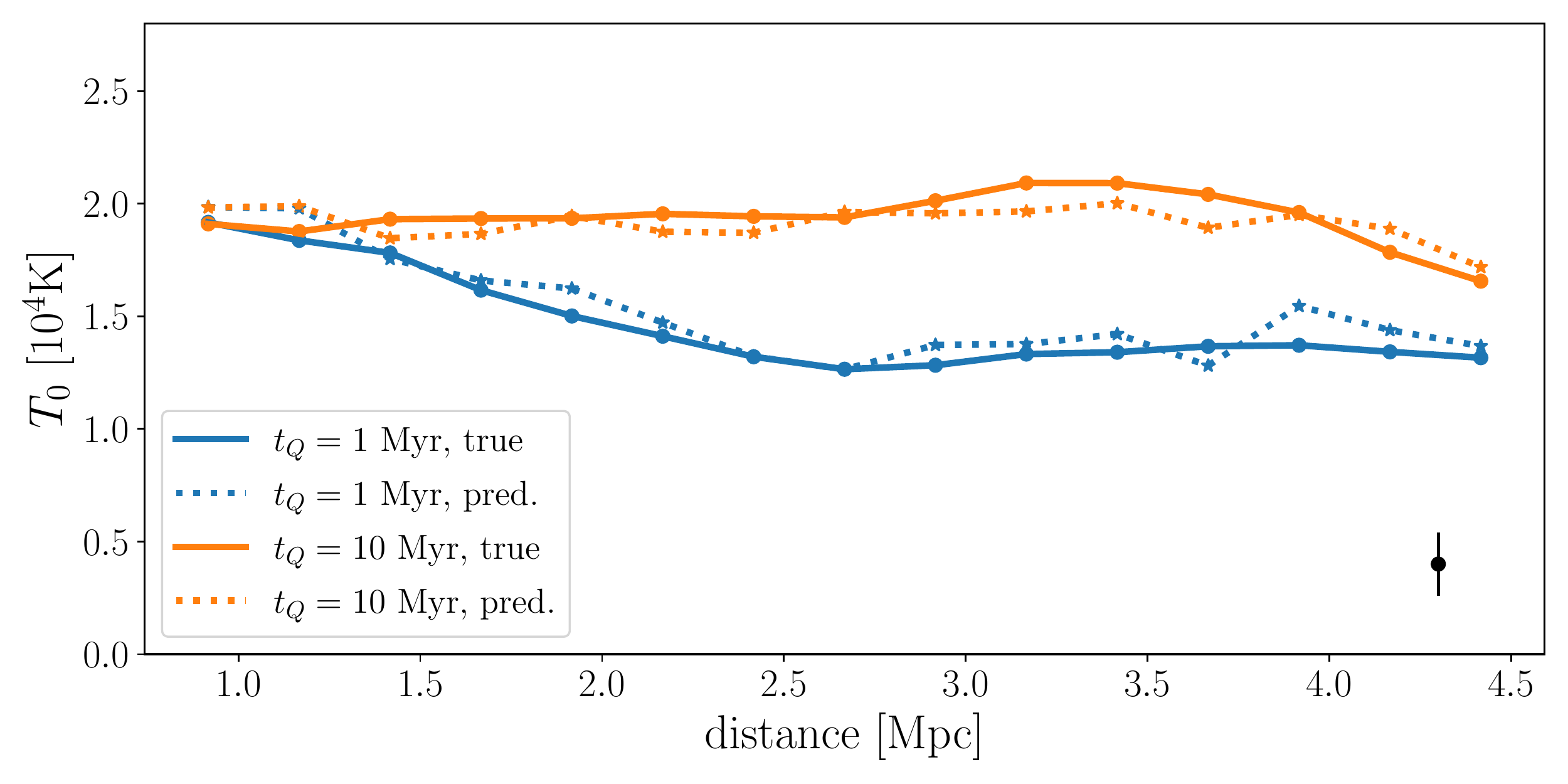}
    \caption{Left: testing result on the most massive halos in our simulations with the model trained only on lower mass halos (similar to Figure \ref{fig:cnnRes}). Right: an example of predicted $T_0$ profile vs the truth. The errorbar in the lower right corner indicates the typical error, $\rm \sqrt{\rm MSE}=1400$ K.  }
    \label{fig:split_halo}
\end{figure*}

\section{Robustness and  generality of the CNN }

In this section, we explore the robustness and generality of the CNN by varying some physical quantities.

\subsection{Robustness against halo masses}

It is still uncertain how massive the host halos of the first quasars are.
Therefore, it is useful to check if the CNN model is robust against the choice of the quasar halo masses. To this end, we train the CNN only with sightlines drawn from less massive halos and test the model on more massive ones. 
We thus sort the sightlines with their halo masses and split them into training, validation and test set with the same ratio of 7:2:1 as in the last section.
The training set contains 57 halos with dark matter halo masses between $1.5\times10^{11}\Msun-2.5\times 10^{11}\Msun$, while the test set contains 9 halos with dark matter halo masses between $3.9\times10^{11}\Msun-1.3\times 10^{12}\Msun$. We train a CNN model with the same hyper-parameters, and we show the test result in the left panel of Figure \ref{fig:split_halo}. The prediction is still very accurate ($\sqrt{\rm MSE}\approx 1400$K), suggesting that using CNN to predict $T_0$ is robust against the choice of the quasar halo mass. It is not surprising because host halo mass correlated mostly with large scale density field, while temperature information is encoded in small scale spectral features. In the right panel of Figure \ref{fig:split_halo}, we show one example sightline drawn from the most massive halo ($1.3\times 10^{12} \Msun$) in the simulation. The solid lines show the true $T_0$ profiles. The  predicted $T_0$ (dotted lines) recover such true $T_0$ profiles very well, and the recovered $T_0$ profiles can easily differentiate the quasar lifetime between $1$ Myr and $10$ Myr.

% \subsection{(in)dependence on distance}
% One key step in augmenting the data is to normalize the spectra with a baseline model.

% \begin{figure*}
%     \centering
%      \includegraphics[width=0.5\textwidth]{fig/true_dist_dist.pdf}
%     \includegraphics[width=0.45\textwidth]{fig/test_true_dist.pdf}
    
%     % \includegraphics[width=1.\textwidth]{fig/fit_ave_temp_diff_spec.pdf}
%     \caption{Left: true distribution of $T_0$ in our data set in two different regions: inner region of $0.5-2.5$ pMpc in blue and outer region of $2.5-4.5$ pMpc in red. Right: testing results for the CNN model trained only on the inner region data ($0.5-2.5$ pMpc). Blue contours ($68\%$ and $95\%$) show the testing result on the same distance range, while red contours show the testing result on the outer region of $2.5-4.5$ pMpc.}
%     \label{fig:diffSP}
% \end{figure*}

\begin{figure*}
    \centering
     \includegraphics[width=\textwidth]{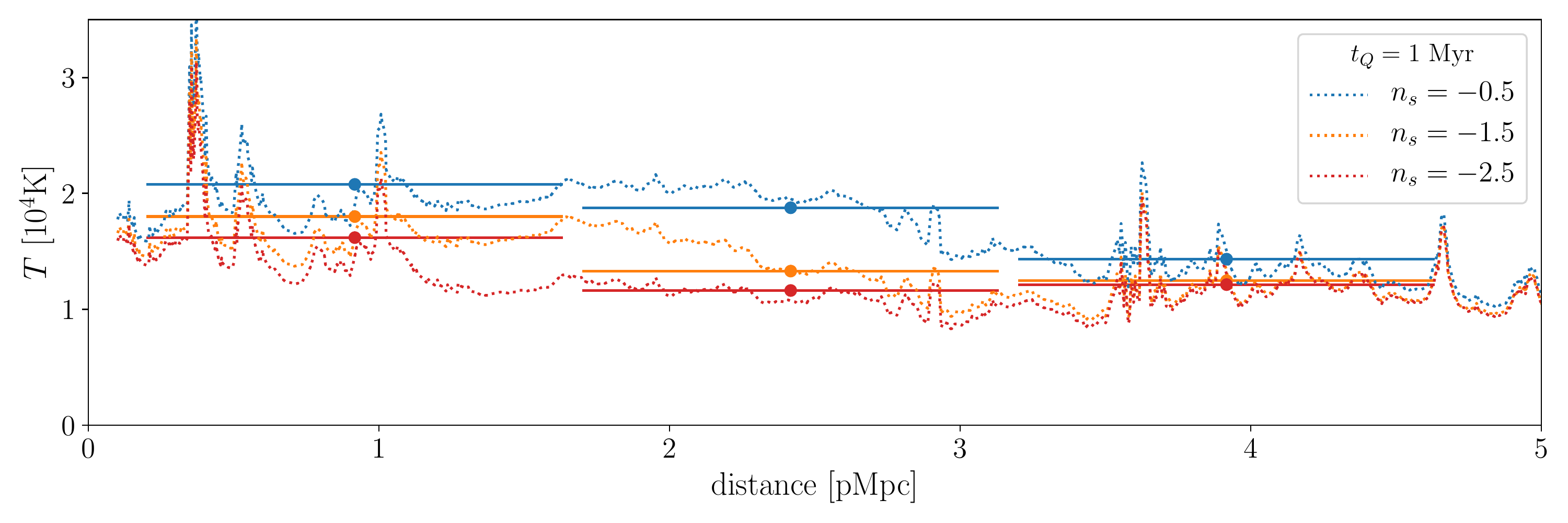}
    \caption{Temperature profile of an example sightline post-processed by three different spectra with the spectral index $n_s=-0.5$ (blue), $-1.5$ (orange) and $-2.5$ (red), respectively. Similar to Figure \ref{fig:T_measurement}, the dotted lines are the temperature profile from the simulation, and the dots are $T_0$ measured with all the pixels marked by each corresponding  horizontal line. The quasar lifetime at this time is $t_Q=1$ Myr.}
    \label{fig:diff_temp}
\end{figure*}

\begin{figure}
    \centering
    \includegraphics[width=0.45\textwidth]{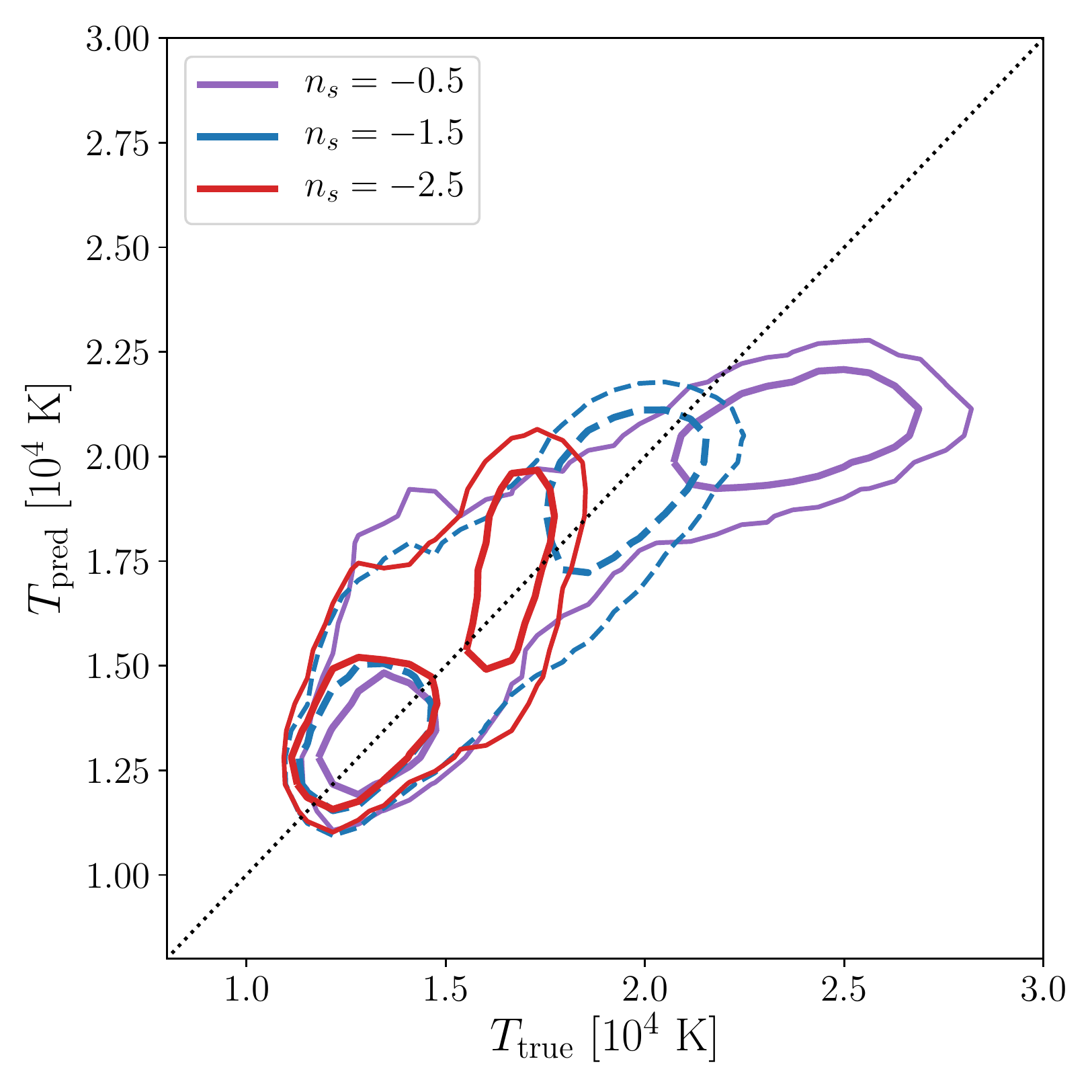}
    \caption{An analog of Figure \ref{fig:cnnRes} (blue dashed contours for $68\%$ and $95\%$ of all data), but now also showing contours for the predicted $T_0$ with CNN trained on $n_s=-1.5$ data but applied to the $n_s=-0.5$ and $n_s=-2.5$ data sets (purple and red contours respectively). The low temperature peaks are predicted well but the high temperature peaks are distorted towards the typical value of the $n_s=-1.5$ case.}
    \label{fig:train15test0525}
\end{figure}

\begin{figure}
    \centering
    \includegraphics[width=0.5\textwidth]{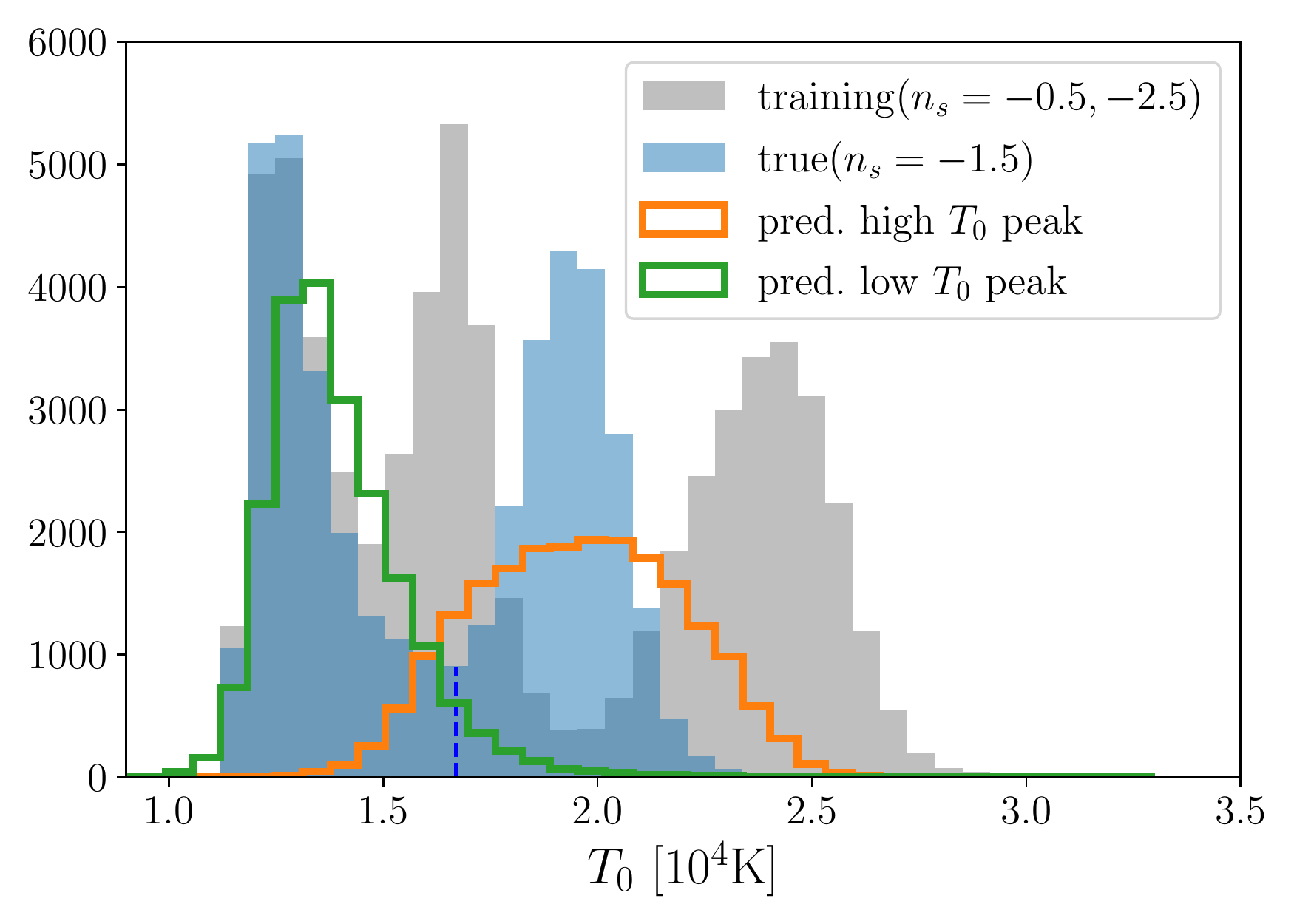}
    \caption{Actual $T_0$ distribution of the training set consisting of sightlines post-processed with the quasar spectral indices of $n_s=-0.5$ and $n_s=-2.5$ (grey filled histogram). Blue filled histogram shows the true $T_0$ distribution of the test set with a different quasar spectral index of $n_s=-1.5$. The vertical dashed blue line shows how we split the lower and higher $T_0$ peaks. Green and orange histograms are predicted $T_0$ distributions of the lower and higher $T_0$ peaks, respectively.}
    \label{fig:traindist}
\end{figure}

\begin{figure}
    \centering
    \includegraphics[width=0.45\textwidth]{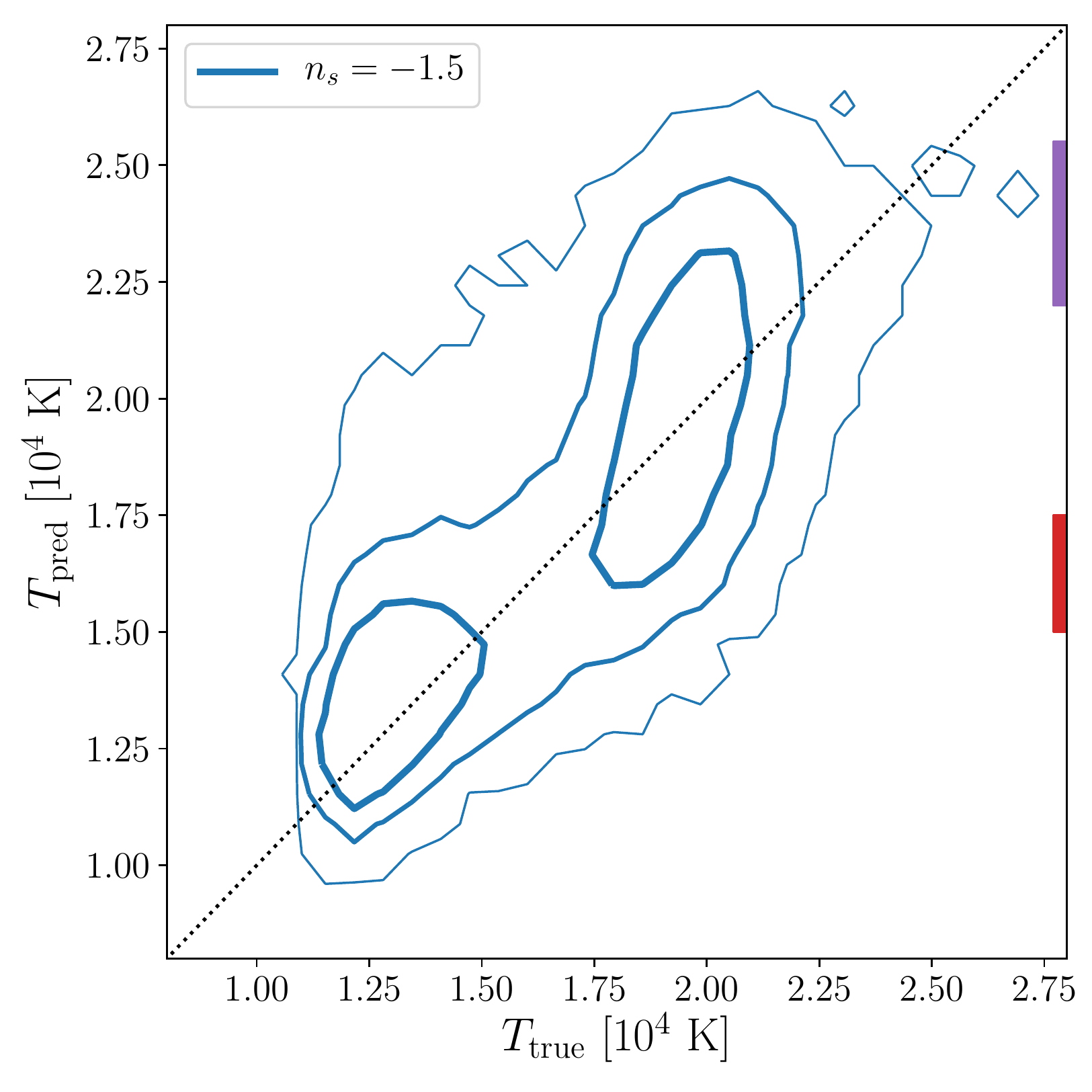}
    \caption{Scatter plot of true $T_0$ vs $T_0$ as predicted by the CNNT for the $n_s=-1.5$ dataset, with the CNN model trained only on $n_s=-0.5$ and $n_s=-2.5$ datasets. The typical ranges of the higher $T_0$ peaks of the training set are indicated by the colored bands on the right.}
    \label{fig:resdiffspec}
\end{figure}

\begin{figure*}
    \centering
    \includegraphics[width=0.32\textwidth]{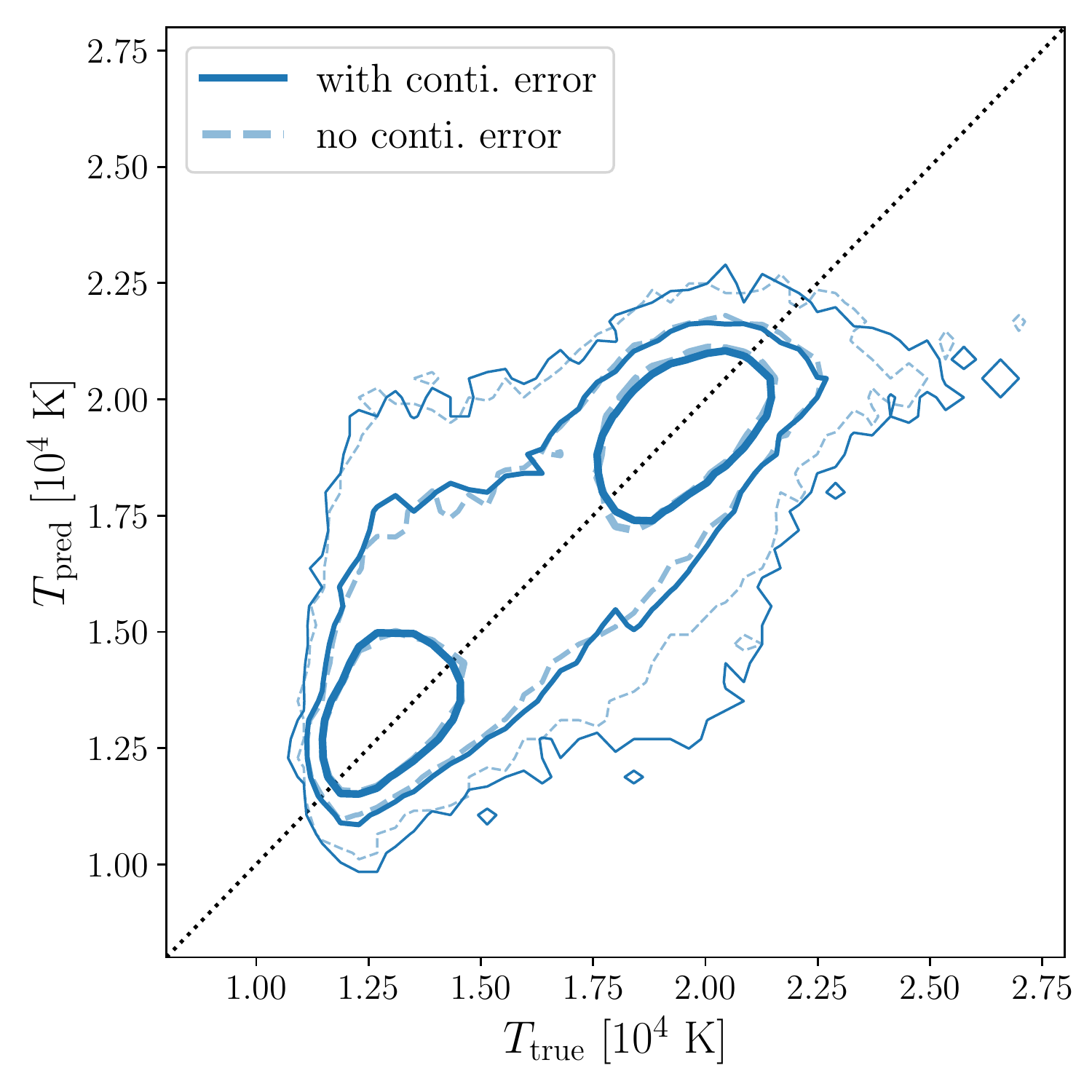}
        \includegraphics[width=0.32\textwidth]{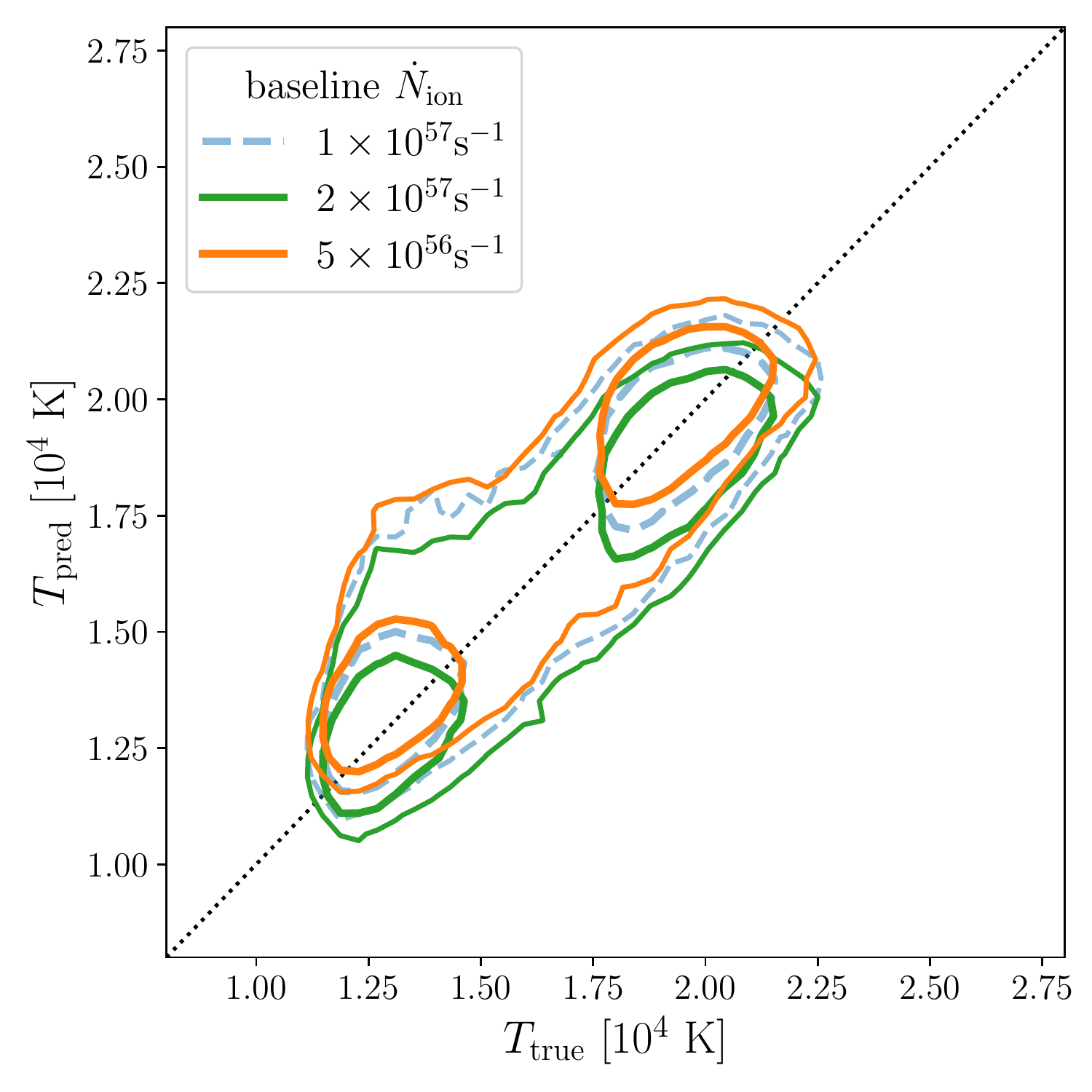}
                \includegraphics[width=0.32\textwidth]{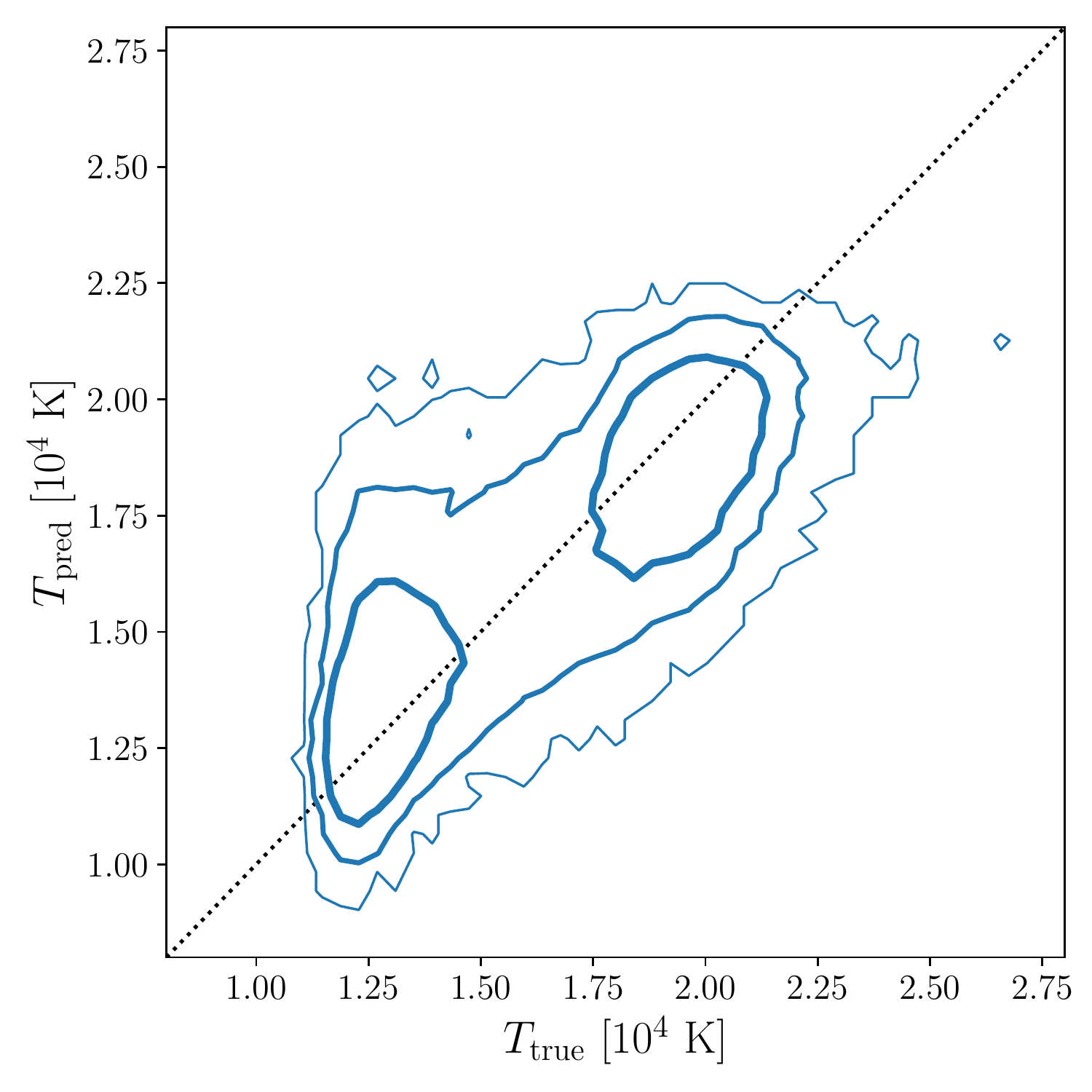}
    \caption{An analog of Figure \ref{fig:cnnRes} for several observational uncertainties. Left: effect of continuum errors in the model (solid contours) vs the fiducial model with the exact continuun (dashed lines). Middle: effect of the mis-estimated quasar emissivity. Blue dashed lines are the fiducial model (the same as the left panel, with the $99.7\%$ contour omitted here for clarity), the green contours show the case with twice overestimated quasar emissivity, while the orange contours show the case with twice underestimated quasar emissivity. Right: effect of saturated pixels, with the CNN model trained on the data with some of the pixels being saturated.}
    \label{fig:uncer}
\end{figure*}

\subsection{Uncertainty in spectral indices} \label{sec:diff_spec}

The spectral index of high-redshift quasars has a large scatter \citep{lusso2015}. The amount of photo-heating depends primarily on the shape of the quasar spectrum, so the variation in the spectral index results in the variation in the IGM temperature behind the HeII I-front \citep{abel1999}. In this section, we consider two extreme cases, a very hard spectrum with $n_s=-0.5$ and a very soft spectrum with $n_s=-2.5$.  We use the same method described in Section \ref{sec:simulations} to produce the spectra, except changing the quasar spectral index to $n_s=-0.5$ and $n_s=-2.5$. In Figure \ref{fig:diff_temp}, we show the temperature profiles of a sightline post-processed by quasars of different spectral indices at $t_Q=1$ Myr. For $n_s=-0.5$ the gas is heated to significantly higher temperature than for the case of the very soft $n_s=-2.5$ spectrum.

The training data produced with a fixed value for the quasar spectral index exhibits a specific value for $T_0$ behind the I-front. Therefore, we do not expect the model trained on such data to accurately predict the temperature if the quasar has a different spectral index. Nevertheless, it is instructive to check how the $T_0$ predicted by such a model deviates from the true one. We thus directly apply the model from Section \ref{sec:cnn} to the datasets with $n_s=-0.5$ and $n_s=-2.5$.  In Figure \ref{fig:train15test0525}, we show the true $T_0$ versus $T_0$ predicted by the same model from Section{sec:cnn}, but applied to datasets with different values of $n_s$. Here for a better comparison, we make the number of data points inside and outside the HeIII I-front roughly the same for each dataset. We find that for data set with $n_s=-0.5$, 
the temperature of the cooler gas (outside the HeIII region) is predicted accurately, but the temperature of the hotter gas (inside the HeIII region) is under-predicted. The median of the predicted temperature of such gas is $\approx 2\times 10^4$ K, similar to the highest temperature end of the training data. This is understandable, because the training set ($n_s=-1.5$) does not have values approaching the true temperature of the gas inside the HeIII region in the $n_s=-0.5$ data set.
However, the two peaks are still clearly differentiated by the CNN trained on such a different data set. On the contrary, for the dataset with $n_s=-2.5$, the temperature of the hotter gas (inside the HeIII region) is over-predicted, with the median value similar to the lower end of the temperature distribution for the gas inside the HeIII region for $n_s=-1.5$. Even though intrinsically the two true $T_0$ peaks are close to each other, the CNN could still do a fair job in classifying both peaks.

We further examine a reverse case -- training the CNN only on the data with $n_s=-0.5$ and $n_s=-2.5$ and testing it on the data with $n_s=-1.5$. From such a training set, we obtain a tri-modal true $T_0$ distribution (see the grey histogram in Figure \ref{fig:traindist}). The true $T_0$ distribution of the test set is plotted in blue, with the typical value of the hotter peak locating in the ``valley'' of the $T_0$ histogram of the training data. We expect the lower $T_0$ peak to be predicted well by the CNN, since there are $T_0$ of similar values in the training data, but it is not guaranteed to be the case for the higher $T_0$ peak ($T_0\sim 2\times10^4$K) -- the training data does not contain many segments with this temperature, and the CNN must interpolate between the two peaks at ($T_0\sim 1.7\times10^4$K) and ($T_0\sim 2.4\times10^4$K) for making the temperature prediction for $T_0\sim 2\times10^4$K. As before, we train the model for $2000$ epochs and compare the predicted results. In Figure \ref{fig:traindist} we plot the predicted $T_0$ distribution for the lower $T_0$ peak in green and higher $T_0$ peak in orange. The lower and higher $T_0$ peak are split at $1.65\times 10^4$ K, corresponding to the ``valley'' of the true $T_0$ distribution in the $n_s=-1.5$ dataset. The predicted lower $T_0$ peak (green) has a similar shape as the true one with a slightly larger width, which is expected since any model always has some scatter. The higher $T_0$ peak (orange) has a much larger width than the true one, but the position of the peak is still at the same place as the true one, i.e.\ the CNN prediction is unbiased. This indicates that even if we do not provide training data of the exact $T_0$, the CNN is able to do some ``interpolation''. In Figure \ref{fig:resdiffspec}, we plot the true $T_0$ versus the predicted $T_0$ of the test set. The $68\%$ contour of the higher $T_0$ one is tilted but confined within the typical value of the hotter gas in the $n_s=-0.5$ and $n_s=-2.5$ datasets, marked by the purple and red color strips on the right.

This test shows that the CNN has a good generalization ability -- it could give a sensible prediction for data which it is not trained for, similar to doing an interpolation. This suggests that we could generalize the CNN to accurately predict $T_0$ for quasars with different spectral indices with a training set with only a few moderately sampled $n_s$.

\subsection{Robustness against uncertainties in continuum fitting, ionizing flux and saturation}

Real observational data have many uncertainties. Notably, the uncertainties from quasar continuum fitting and the estimation of total ionizing flux can be $\gtrsim 10\%$ for individual quasars \citep{chen2022b}. Also, there are inevitably some saturated regions inside proximity zones due to the finite signal-to-noise ratio. Although rigorously analyzing all observation uncertainties is beyond the scope of this paper, we estimate potential errors due to these three observational uncertainties separately.

We first produce a test set to mimic data with continuum errors. This set is produced in a similar way to Figure 9 in \citet{chen2021b}. Namely, we draw a continuum with error linearly dependent on distance:
$$ C(d)=\tilde{C}(d)*(1+\alpha d)$$,
where $\tilde{C}$ is the true continuum, $d$ is the distance from the quasar, and $\alpha$ is a random number drawn from a Gaussian distribution with standard deviation of $0.05$/pMpc. We directly apply the CNN model in Section \ref{sec:cnn} to this test set. We calculate the $\sqrt{\rm MSE}$ for this test set the same way as in Section \ref{sec:cnn}. We find that the $\sqrt{\rm MSE}\approx 1450$ K, similar to the case without continuum errors. This insensitivity to continuum errors is most likely because the features extracted by the CNN are all on small scales. The continuum errors manifest themselves as large-scale bias, while significantly impacting the Doppler broadening of small-scale features. In the left panel of Figure \ref{fig:uncer} we show true $T_0$ versus predicted $T_0$ for the test data with continuum errors in solid blue contours. Compared with test data without continuum errors (dashed contours), the accuracy of the CNN prediction is not significantly degraded.

In real observations we cannot precisely measure the ionizing emissivity $\dot{N}_{\rm ion}$ of a quasar, and this uncertainty introduces an error in the baseline $\tilde{\tau}$ in Equation (\ref{eq:Rtau}), which, in turn, eventually propagates into the predicted $T_0$. To estimate such an error, we recalculate $R_{\tau}$ using $\dot{N}_{\rm ion}$ which is over- and under-estimated by a factor of 2: instead of calculating the baseline $\tilde{\tau}$ with the correct value of $\dot{N}_{\rm ion}=1\times 10^{57} \rm s^{-1}$, we calculate it with $\dot{N}_{\rm ion}=2\times 10^{57} \rm s^{-1}$ and $\dot{N}_{\rm ion}=5\times 10^{56} \rm s^{-1}$, respectively. Then we directly apply the same CNN model from Section \ref{sec:cnn} (i.e.\ trained with $\dot{N}_{\rm ion}=1\times 10^{57} \rm s^{-1}$) to predict $T_0$ of these two datasets. In the middle panel of Figure \ref{fig:uncer}, we show the $68\%$ and $95\%$ contours of the results. We find that there is a small shift in the pattern compared to the blue dashed contours, which show the results without $\dot{N}_{\rm ion}$ error. This shift suggests that the error in $\dot{N}_{\rm ion}$ can be captured by a bias term in the CNN. However, even applying the CNN directly will not introduce a significant error in the predicted $T_0$ -- a factor of 2 error in $\dot{N}_{\rm ion}$ only results in a 3\% error in $T_0$.

Finally, we test how saturated regions impact $T_0$ prediction. In reality, because of the limited signal-to-noise, regions of higher density, such as cosmic filaments, result in saturated absorption (pixel values indistinguishable from noise). Therefore, to estimate the error due to such saturation, we first mimic the saturation in the data by capping the maximum $\log_{10}R_\tau$ to be $0.5$. This level of saturation is typical of existing observational data at $\gtrsim 3$ pMpc from the quasar \citep{chen2021b}.  By putting an upper limit on $\log_{10}R_\tau$, we significantly change the small scale features in such regions. Therefore, directly applying the CNN model in Section \ref{sec:cnn} does not work well in predicting $T_0$. Instead, we re-train the CNN model with the same architecture but using the training and validation data also capped \ref{sec:cnn} at $\log_{10}R_\tau=0.5$. We find that the validation loss degrades to $\sqrt{\rm MSE} \approx 1700$ K from $\sqrt{\rm MSE} \approx 1400$ K in Section \ref{sec:cnn}). In the right panel of Figure \ref{fig:uncer}, we show the result of the test set. We find that the contours deviate  more from the perfect diagonal line than previous cases. The two $68\%$ peaks can still be separated by a horizontal line but barely. This test suggests that high signal-to-noise spectra are needed for measuring $T_0$ and distinguishing the gas inside and outside the HeIII region by its temperature.

%The experiments in this subsection shows that the CNN is robust against errors in continuum estimation and the intrinsic ionizing flux of the quasar. However, saturation is a bigger concern in CNN -- after all, saturation means the loss of small scale features from which $T_0$ information is extracted. Our test indicates that at distance $> 3$ pMpc from the quasar, recovering $T_0$ accurately would be increasingly hard with the simple CNN presented in this paper and may require more complex neural architecture. NG: belongs to conclusions

\section{Conclusions}

In this work we experiment with using a CNN to reconstruct the profile of the temperature at the mean cosmic density, $T_0$, in the proximity zone of $z\approx 6$ quasars. We use a baseline model to normalize the optical depth to reduce the distance dependency of the quasar ionization flux in the proximity zone. By doing so we produce an augmented dataset.
We find that a simple CNN can predict the temperature to an accuracy of around $7-10\%$. We test the CNN performance on different quasar host halo masses between $10^{11} \Msun \sim 10^{12} \Msun$ and find that the CNN is robust against the uncertainty in the quasar halo mass in this range. We also explore the generality of the CNN for different quasar spectra and find that even if the training set does not include data with the correct quasar spectrum (i.e.\ with the correct post-HeIII-front temperature), the CNN can still find the correct range of $T_0$, showing good generality. 

We also consider how observational uncertainties in quasar continuum fitting, ionizing flux and saturated pixels impact the goodness of prediction. Our experiments show that the CNN is robust against errors in the continuum estimation and in the intrinsic ionizing flux of the quasar. However, pixel saturation due to the finite signal-to-noise ratio is a bigger concern for the CNN -- after all, saturation means the loss of small scale features from which $T_0$ information is extracted. Our test indicates that at distance $> 3$ pMpc from the quasar, recovering $T_0$ accurately would be increasingly hard with the simple CNN presented in this paper and may require more complex neural architecture.

With this method we could differentiate gas inside and outside the HeIII region created by the quasar. Because the size of the HeIII region is closely related to the total lifetime of the quasar, this method could help us constrain the quasar lifetime. Specifically, our method works on $\sim $ Mpc scales, corresponding to a total quasar lifetime of $\sim$ Myr scales. This method can also be extended to quasars at lower redshifts before global HeII reionization ($z\gtrsim 4$). In the future, we would also examine the uncertainties from observations more rigorously and apply this method to real observational data. 
With thirty-meter-class telescopes coming online in the next decade, we could expect hundreds or even thousands of high resolution, high signal-to-noise spectra taken from such high-$z$ quasars.
Since quasar lifetime is a critical quantity related to their growth yet has not been constrained accurately, IGM temperature reconstruction in their proximity zones using CNN can be a promising way forward and can further shed light on the long-standing mystery of how their central supermassive black holes grow.

% The last numbered section should briefly summarise what has been done, and describe
% the final conclusions which the authors draw from their work.

% \section*{Acknowledgements}

% The Acknowledgements section is not numbered. Here you can thank helpful
% colleagues, acknowledge funding agencies, telescopes and facilities used etc.
% Try to keep it short.

%%%%%%%%%%%%%%%%%%%%%%%%%%%%%%%%%%%%%%%%%%%%%%%%%%
% \section*{Data Availability}

% The inclusion of a Data Availability Statement is a requirement for articles published in MNRAS. Data Availability Statements provide a standardised format for readers to understand the availability of data underlying the research results described in the article. The statement may refer to original data generated in the course of the study or to third-party data analysed in the article. The statement should describe and provide means of access, where possible, by linking to the data or providing the required accession numbers for the relevant databases or DOIs.

%%%%%%%%%%%%%%%%%%%% REFERENCES %%%%%%%%%%%%%%%%%%

% The best way to enter references is to use BibTeX:

\bibliographystyle{mnras}
\bibliography{ms} % if your bibtex file is called example.bib

% Alternatively you could enter them by hand, like this:
% This method is tedious and prone to error if you have lots of references
%\begin{thebibliography}{99}
%\bibitem[\protect\citeauthoryear{Author}{2012}]{Author2012}
%Author A.~N., 2013, Journal of Improbable Astronomy, 1, 1
%\bibitem[\protect\citeauthoryear{Others}{2013}]{Others2013}
%Others S., 2012, Journal of Interesting Stuff, 17, 198
%\end{thebibliography}

%%%%%%%%%%%%%%%%%%%%%%%%%%%%%%%%%%%%%%%%%%%%%%%%%%

%%%%%%%%%%%%%%%%% APPENDICES %%%%%%%%%%%%%%%%%%%%%

% \appendix

% \section{Some extra material}
% \begin{figure*}
%     \centering
%     \includegraphics[width=\textwidth]{fig/ave_temp_prof_diff_spec_01Myr.pdf}
%      \includegraphics[width=\textwidth]{fig/ave_temp_prof_diff_spec_1Myr.pdf}
%     % \includegraphics[width=1.\textwidth]{fig/fit_ave_temp_diff_spec.pdf}
%     \caption{}
%     \label{fig:diffSP}
% \end{figure*}

% \begin{figure*}
%     \centering
%     \includegraphics[width=0.8\textwidth]{fig/2_diffSP.pdf}
%     \caption{Similar to Figure \ref{fig:exampleLOS}, but different spectral indices at $t_Q=10$ Myr. The harder the spectrum, the higher the temperature inside the HeIII region. Because the number of photons above $13.6$ eV is fixed in our simulation, a harder spectrum results in a larger HeIII region at a given quasar lifetime.}
%     \label{fig:tQ1Myr}
% \end{figure*}

% If you want to present additional material which would interrupt the flow of the main paper,
% it can be placed in an Appendix which appears after the list of references.

%%%%%%%%%%%%%%%%%%%%%%%%%%%%%%%%%%%%%%%%%%%%%%%%%%

% Don't change these lines
\bsp	% typesetting comment
\label{lastpage}
\end{document}